\documentclass[%
 reprint,
 amsmath,amssymb,
 aps,
 prl,
]{revtex4-2}

\usepackage{graphicx}

\usepackage[utf8]{inputenc}
\usepackage{pgfplots}
\usepgfplotslibrary{groupplots,dateplot}
\usetikzlibrary{patterns,shapes.arrows}
\pgfplotsset{compat=newest}

\usepackage{dcolumn}
\usepackage{bm}
\usepackage{hyperref}
\usepackage{xcolor}
\usepackage{ulem}

\begin{document}

\preprint{APS/123-QED}

\title{Cellular gradient flow structure connects \\ single-cell-level rules and population-level dynamics}

\author{Shuhei A. Horiguchi}
\author{Tetsuya J. Kobayashi}
    \altaffiliation[Also at ]{Institute of Industrial Science, The University of Tokyo}
 \email{tetsuya@mail.crmind.net}
\affiliation{
Graduate School of Information Science and Technology, The University of Tokyo
}

\date{\today}

\begin{abstract}
In multicellular systems, the single-cell behaviors should be coordinated consistently with the overall population dynamics and functions.  
However, the interrelation between single-cell rules and the population-level goal is still elusive.
In this work, we reveal that these two levels are naturally connected via a gradient flow structure of the heterogeneous cellular population and that biologically prevalent single-cell rules such as unidirectional type-switching and hierarchical order in types emerge from this structure.
We also demonstrate the gradient flow structure in a standard model of the T-cell immune response. This theoretical framework works as a basis for understanding multicellular dynamics and functions.
\end{abstract}

\maketitle


\section{Introduction}

Multicellular systems are organized dynamically and robustly to shape various populational patterns required to achieve biological functions \cite{Alenzi2009,Lee2017,VonDassow2000}. 
Because the population dynamics is realized by single-cell-level processes, i.e., cellular proliferation, death, migration, and differentiation, behaviors of individual cells should be coordinated consistently with the overall population dynamics, which may rule the single-cell processes. 

For example, phenotypic switching and differentiation of a cell in a population are often unidirectional. Moreover, the multiple cell types are hierarchically ordered, and their kinetic properties, e.g., type-switching and proliferation rates, also seem to be coupled. The existence of hierarchy, or equivalently, acyclic cell-type lineage structures and kinetic coupling are prevalent in multicellular systems from immunity to development \cite{Wagner2018,Cheng2020,Mani2021,Ruijtenberg2016}.
These single-cell rules emerging in a population may be related to the functions and the coordination of the population.

The notion of the epigenetic landscape has been used pervasively to describe the directional dynamics among hierarchical cellular types, where individual cells are likened to balls rolling down the landscape \cite{Waddington1957,Kauffman1969,Wang2008}.
However, the interrelation of such single-cell-level dynamics and landscape with the overall population dynamics and their function has been rarely investigated except in a few prescient works pointing out its importance \cite{Furusawa1998,Furusawa2001,Furusawa2012}. 

In this work, we show that the interrelation between single-cell and population levels can emerge from a gradient flow structure of a population.
We model the desirable population distribution for achieving a biological function by the landscape of a utility function.
Then, we derive the population dynamics that maximizes the utility given biological costs of single-cell processes, which results in a gradient flow of the utility function.
We demonstrate as an example that the standard model of T-cell population dynamics in the acute immune response \cite{DEBOER2013} can be understood as the gradient flow.
From the populational gradient flow structure, the single-cell-level landscape emerges, from which the unidirectional type switching, hierarchical cell-type, and kinetic couplings are generally derived.
Moreover, the single-cell landscape is related to the population-level utility landscape as its functional variation.
Our result can work as a theoretical basis to bridge single-cell-level rules and behaviors with population-level dynamics and functions of multicellular systems.

\section{Gradient flow of cellular population dynamics}
We firstly introduce the governing equations of the heterogeneous cellular population dynamics. 
We consider a large heterogeneous population of cells with different types.
The state of population at time $t$ is characterized by $n_t = \{n_t(x)\}_{x\in X}$, where $n_t(x) \geq 0$ is the population size of the cells with type $x\in X$. 
Here $X$ is a set of all possible types.
The state of the population changes over time by the following cellular actions: growth (proliferation minus death), type switching, and immigration (recruitment) of new cells from outside of the population. 
Their rates are assumed to be type-dependent such that $g_t(x)$ is the growth rate of type $x$, $v_t(x,y)\geq 0$ is the type-switching rate from type $x$ to type $y$, and $m_t(x) \geq 0$ is the immigration rate of type $x$.
In this work, we focus on the case where $X$ is discrete, but it can be easily extended to the continuum case.
Then, the population dynamics of the cells can be described by the following equations
\begin{equation}\label{eq:continuity}
	\begin{aligned}
\frac{dn _{t}( x)}{dt} 
	= &n _{t}( x) g_{t}( x) +m_t( x) \\
	&-\sum _{y \in X}( n _{t}( x) v_{t}( x,y) -n _{t}( y) v_{t}( y,x)) \\
	=: & F_x(n, g,m,v),		
	\end{aligned} 
\end{equation}
where we approximate $n_t(x)$ as a continuous variable, which is valid if the population size is large enough. 

Next, we connect the population dynamics with biological functions. 
To this end, we introduce a \textit{utility function} $U_t(n)$.
The utility function $U_t(n)$ abstractly represents how good a given population distribution $n$ is under the situation at time $t$.
The time dependence of $U_t(n)$ is essential for modeling various biological situations.
For example, if a pathogen invades our body, a particular population distribution $n$ of immune cells would work more effectively than another $n'$, which is represented as $U_t(n)>U_t(n')$. 
The utility may change after the eviction of the pathogen, which is captured by the time dependence of $U_t(n)$.
Another example of the utility function is for developmental processes. A specific pattern of differentiated cells would be required at time $t$, which may change as the development progresses.
Therefore, the dynamics that can induce the population distribution $n_{t}(x)$ into the one with a higher utility more promptly would be more functional than other dynamics.

However, the rates of growth, immigration, and type switching cannot be arbitrarily high due to the biological cost of those processes and physical constraints.
In order to account for it, we introduce the \textit{cost function} $C_n(g,m,v)$. 
The cost function abstractly characterizes the instantaneous biological cost of taking cellular actions at given rates $(g,m,v)$, which is nonnegative and dependent on the current population $n$.
In this work, we consider the cost function to have the following form
\begin{equation}\label{eq:cost}
	\begin{aligned}
	C_n(g,m,v) 
    = & \frac{1}{2}\sum_{x \in X} n(x) w_g(x) g(x)^2\\
    &+ \frac{1}{2}\sum_{x \in X} w_m(x) m(x)^2 \\
    &+ \frac{1}{2}\sum_{x, y\in X} n(x) w_v(x,y) v(x,y)^2 \geq 0,
	\end{aligned}
\end{equation}
where $w_g$, $w_m$, and $w_v$ are positive weights whose values depend on the single-cell level mechanisms of actions.

This form of the cost function is derived from three assumptions: 1) costs for different cellular actions are independent, i.e., the total cost is just a sum of them, 2) growth costs and type-switching costs are proportional to the current cell number, and 3) for each cellular action, the cost is a strictly convex smooth function of the rate.
The second assumption is reasonable because growth and type-switching costs are incurred for each cell in the current population.
In contrast, since immigration is usually independent of the current population, the immigration cost is not proportional to the current cell number.
The third assumption is crucial to prohibit the optimal action rates from being arbitrarily high. 
Such unrealistic behavior can happen if the cost grows slowly as the rates increase, and the utility can cancel it out. 
While we focus here on the quadratic cost function, the simplest convex function, our theory can be extended to more general convex functions.

We consider the population dynamics of Eq.~\eqref{eq:continuity} where the rates $(g_t, m_t, v_t)$ are determined to maximize the value of the utility function under the cost.
\begin{equation}\label{eq:maximization_prob}
	\underset{g,m,v}{\text{maximize}} ~~ \mathrm{Diff}_{n_t} U_t(g,m,v) - C_{n_t} (g,m,v),
\end{equation}
where $\mathrm{Diff}_{n_t} U_t(g,m,v)$ is the time derivative of $U_t$ through the time evolution of $n_t$ given rates $(g,m,v)$:
\begin{equation*}
    \mathrm{Diff}_{n_t} U_t(g,m,v) 
	:= \sum_{x\in X} \frac{\delta U_t(n_t)}{\delta n}(x) F_x(n_t, g,m,v)
\end{equation*}

We could obtain the explicit form of the unique optimum $(g_t^\dagger, m_t^\dagger, v_t^\dagger)$ of the above optimization problem as follows (see Appendix for derivation)
\begin{subequations}
\label{eq:optimal_rates}
\begin{align}
	g_t^\dagger(x) &= \frac{1}{w_g(x)}\frac{\delta U_t(n_t)}{\delta n}(x),\label{eq:optimal_growth_rate}\\
	m_t^\dagger(x) &= \frac{1}{w_m(x)} \left[ \frac{\delta U_t(n_t)}{\delta n}(x) \right]_{+},\label{eq:optimal_immigration_rate}\\
	v_t^\dagger(x,y) &= \frac{1}{w_v(x,y)} \left[ \overline{\nabla}\frac{\delta U_t(n_t)}{\delta n}(x,y) \right]_{+},\label{eq:optimal_switch_rate}
\end{align}
\end{subequations}
where $[a]_{+}$ is the positive part of $a \in \mathbb{R}$, and $\overline{\nabla}$ is the discrete gradient operator, i.e., for any $\phi:X\rightarrow \mathbb{R}$, $\overline{\nabla}\phi(x,y) := \phi(y) - \phi(x)$.
Note that these optimum rates are scale-invariant: they are invariant under the rescaling of the utility function $U_t$ and the cost function $C$ with the same factor.

Let us consider the dynamics with the optimal rates $(g_t, m_t, v_t)=(g_t^\dagger, m_t^\dagger, v_t^\dagger)$.
Under this dynamics, the value of the utility function $U_t(n)$ evolve as
\begin{align} \label{eq:dU_dt_general}
	\frac{d}{dt}[U_t(n_t)] 
	&= \frac{\partial U_t}{\partial t} (n_t) + \mathrm{Diff}_{n_t} U_t(g^\dagger_t,m^\dagger_t,v^\dagger_t) \nonumber\\
	&= \frac{\partial U_t}{\partial t} (n_t) +2 C_{n_t}(g^\dagger_t, m^\dagger_t, v^\dagger_t).
\end{align}
Since the instantaneous cost $C_{n_t}$ is nonnegative, the value of the utility function always increases when $\frac{\partial U_t}{\partial t}$ vanishes.
Indeed, when $U_t$ does not depend on $t$, the dynamics is a generalized gradient flow of $U$, where $C$ is called as a dissipation function \cite{Ambrosio2005,Mielke2011,Mielke2014}.

\section{T-cell immune response model}

\begin{figure}[htp]
\centering
\includegraphics[width=8cm]{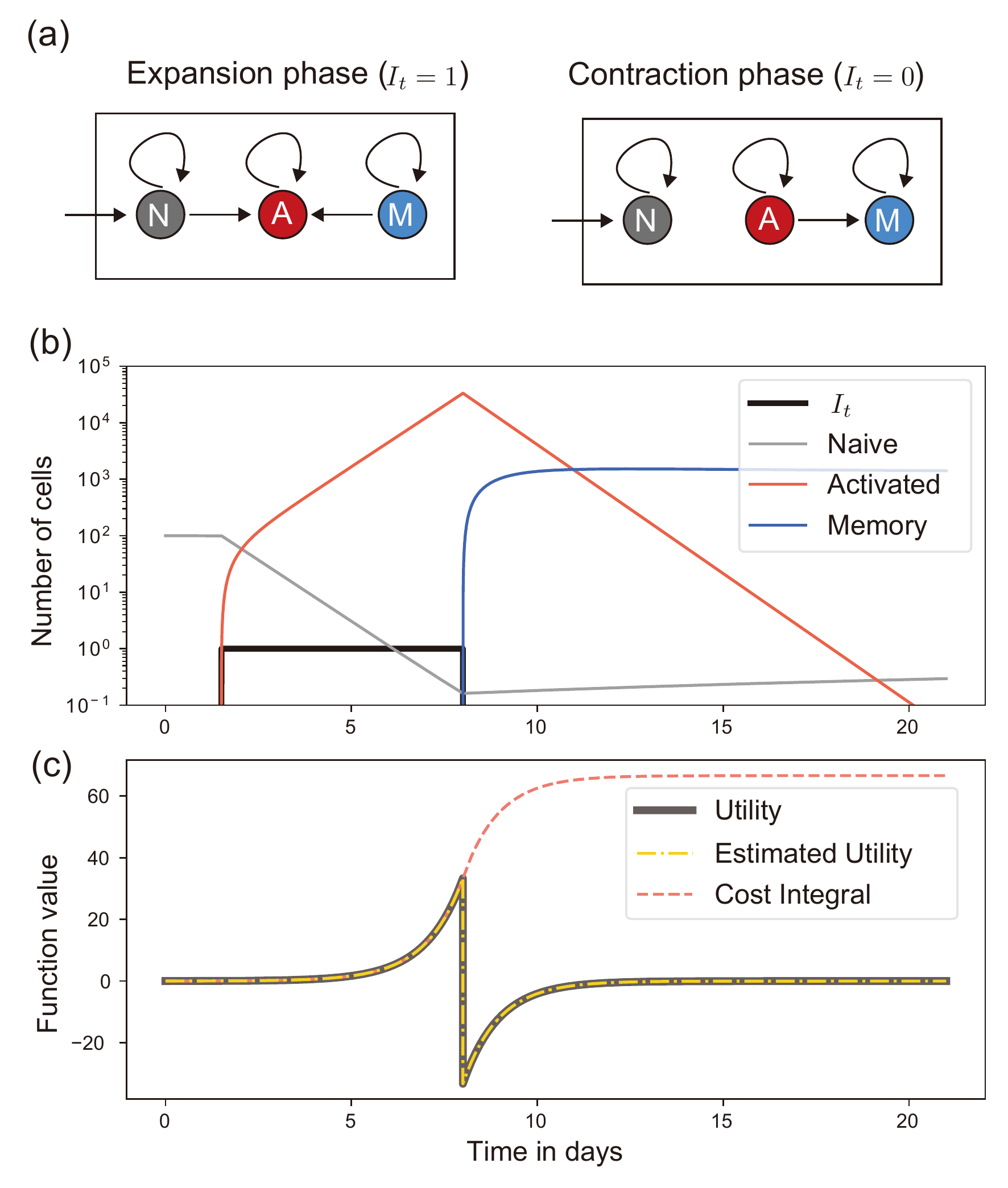}
\caption{(a) Schematic illustration of the immune response model. N: naive, A: activated, M: memory T cells. (b) Time evolution of the numbers of T cells of the three types. (c) Time evolution of the utility $U_t(n_t)$, the estimated utility based on Eq.~\eqref{eq:dU_dt_general}, and the cost integral $2 \int_0^t C_{n_\tau}(g_\tau^\dagger,m_\tau^\dagger,v_\tau^\dagger) d\tau $. The parameters are $m_N=0.01$, $g_N=0.001$, $g_{A0}=-1$, $g_{A1}=2$, $g_M=-0.01$, $v_{N,A}=v_{M,A}=1$, $v_{A,M}=0.05 ~ \mathrm{day}^{-1}$, $w_{0}=10^{-3}, \rho_N= \rho_M=1$. The simulation starts at $t=0$ with $n_0(N)=100,n_0(A)=n_0(M)=0$ and the value of $I_t$ switches at time $\tau_0=1.5$ and $\tau_1=8 ~\mathrm{day}$.}
\label{fig:Tcell_dynamics}
\end{figure}

We demonstrate that a model of T-cell population dynamics in the acute immune response \cite{DEBOER2013} is a gradient flow in our sense. Here we introduce a slightly modified version of the model proposed in \cite{DEBOER2013}. The model assumes three types of T cells, naive ($N$), activated effector ($A$), and memory ($M$), and the numbers of these cells $n_t(N), n_t(A), n_t(M)$ are described by the following ordinary differential equations
\begin{equation}\label{eq:T-cell_model}
    \begin{aligned}
        \frac{dn_t(N)}{dt} &= m_N + g_N n_t(N) - v_{NA}I_tn_t(N), \\
	\frac{dn_t(A)}{dt} &= (g_{A0} + g_{A1} I_t) n_t(A) + v_{NA}I_tn_t(N) \\
	 &\quad\quad + v_{MA}I_tn_t(M) - v_{AM}(1-I_t)n_t(A),\\
	\frac{dn_t(M)}{dt} &= g_Mn_t(M) - v_{MA}I_tn_t(M)\\
	&\quad\quad+ v_{AM}(1-I_t)n_t(A),
    \end{aligned}
\end{equation}
where $g, m, v$ are constant growth, immigration and type-switching rates, and $I_t$ represents the temporal change in the environmental situation such that $I_t =0$ when the immune cells should contract to recover to the normal state, and $I_t=1$ when they should expand to eliminate pathogens (Fig.~\ref{fig:Tcell_dynamics}~(a)). We assume on/off transition:
\begin{equation}\label{eq:on_off_input}
	I_t = \begin{cases}
		1 & \text{if}~\tau_0 \leq t <  \tau_1,\\
		0 & \text{otherwise}.
	\end{cases}
\end{equation}

A typical time evolution is depicted in Fig.~\ref{fig:Tcell_dynamics}~(b). In the expansion phase ($\tau_0 \leq t < \tau_1$), the number of activated T cells rapidly increases, whereas, in the contraction phase ($\tau_1 \leq t$), the number of memory T cells increases instead.

There are several versions of the T-cell immune response model, and their mathematical properties were investigated in \cite{DeBoer1995,Antia2003,Anelone2016}.
The model we introduced in Eqs.~\eqref{eq:T-cell_model}–\eqref{eq:on_off_input} is simple yet includes immunologically realistic factors.
Notably, a simplified version of the model introduced here was shown to reproduce experimental data \cite{DEBOER2013,DeBoer2001}.
In Appendix, we list some of the other T-cell immune response models and discuss their gradient flow structures.

The model Eqs.~\eqref{eq:T-cell_model}–\eqref{eq:on_off_input} is a gradient flow in our framework on three-type space $X=\{ N, A, M \}$ with the following utility function $U_t$ and the cost function $C$. The utility function is a time-dependent linear function
\begin{equation} \label{eq:utility}
	U_t(n) = u_N n(N) + u_A^{(I_t)} n(A) + u_M n(M),
\end{equation}
with coefficients $u_N:=g_N \rho_N w_0$, $u_A^{(I_t)} := (g_{A0} + g_{A1} I_t)w_0$, and $u_M:=g_M \rho_M w_0$, indicating that each cell type has different importance depending on the environmental situation.
The cost function (Eq.~\eqref{eq:cost}) is specified with the weights
\begin{align}\label{eq:weights_T-cell}
	w_{g}(N) &= \rho_{N}w_0,\quad w_g(A) = w_0,\quad w_g(M) = \rho_M w_0,\nonumber\\
	w_{m}(N) &= \frac{u_N}{m_N} = \frac{g_N \rho_N}{m_N} w_0,\\
	w_{v}(N,A) &= \frac{u_A^{(1)}-u_N}{v_{NA}} = \frac{(g_{A0}+g_{A1}) - g_N \rho_N}{v_{NA}}w_0,\nonumber\\
	w_{v}(M,A) &= \frac{u_A^{(1)}-u_M}{v_{MA}} = \frac{(g_{A0}+g_{A1}) - g_M \rho_M}{v_{MA}}w_0,\nonumber\\
	w_{v}(A,M) &= \frac{u_M - u_A^{(0)}}{v_{AM}} = \frac{g_M \rho_M - g_{A0}}{v_{AM}} w_0,\nonumber
\end{align}
and all the other weights are $+\infty$. Here, $w_0$, $\rho_N$, and $\rho_M$ are arbitrary positive constants. By taking account of the scale invariance of the optimal rates, $w_0$ is the scaling factor for the utility function and the cost function. We define it as the same as the growth weight for activated T cells. $\rho_N$ and $\rho_M$ are the relative growth weights for naive and memory T cells. 

To demonstrate that the model is actually a gradient flow, we numerically calculated $U_t(n_t)$, the value of utility function along the time evolution of $n_t$ (Fig.~\ref{fig:Tcell_dynamics}~(c)).
The result shows that $U_t(n_t)$ is monotonically increasing except at the change point $t=\tau_1$, where $\frac{\partial U_t}{\partial t}$ in Eq.~\eqref{eq:dU_dt_general} becomes nonzero. Thus, the gradient flow structure completely explains this behavior.

Finally, we discuss that the gradient flow structure is qualitatively consistent with biologically plausible parameters.
The structure imposes the positivity of the weights (Eqs.~\eqref{eq:weights_T-cell}) of the cost function. Rearranging the terms, we obtain the following constraints on the growth rates:
\begin{equation}\label{eq:ineq_params_Tcell}
    \begin{aligned}
	0 < &g_N  < (g_{A0}+g_{A1}) \frac{1}{\rho_N},\\
	g_{A0}\frac{1}{\rho_M} < &g_M < (g_{A0}+g_{A1})\frac{1}{\rho_M}.
\end{aligned}
\end{equation}
If these inequalities do not hold, the increase in utility is no longer guaranteed.
The existence of expansion and contraction in the immune response naturally requires that the growth rate for the activated T cells in the contraction phase be negative ($g_{A0} < 0$) and the growth rate for the activated T cells in the expansion phase be positive ($g_{A0}+g_{A1} > 0$).
If, in addition, the growth rate $g_N$ for the naive T cells is positive, there exists the weight parameters $\rho_N$ and $\rho_M$ satisfying the constraints.
According to \cite{DENBRABER2012}, naive T cells in humans have a relatively high proliferation rate to maintain the size of the naive T-cell population, implying the growth rate $g_N$ for the naive T cells to be positive.
Thus, the gradient flow structure is consistent with biologically plausible parameters.

\begin {figure}[t]
\centering
\includegraphics[width=8cm]{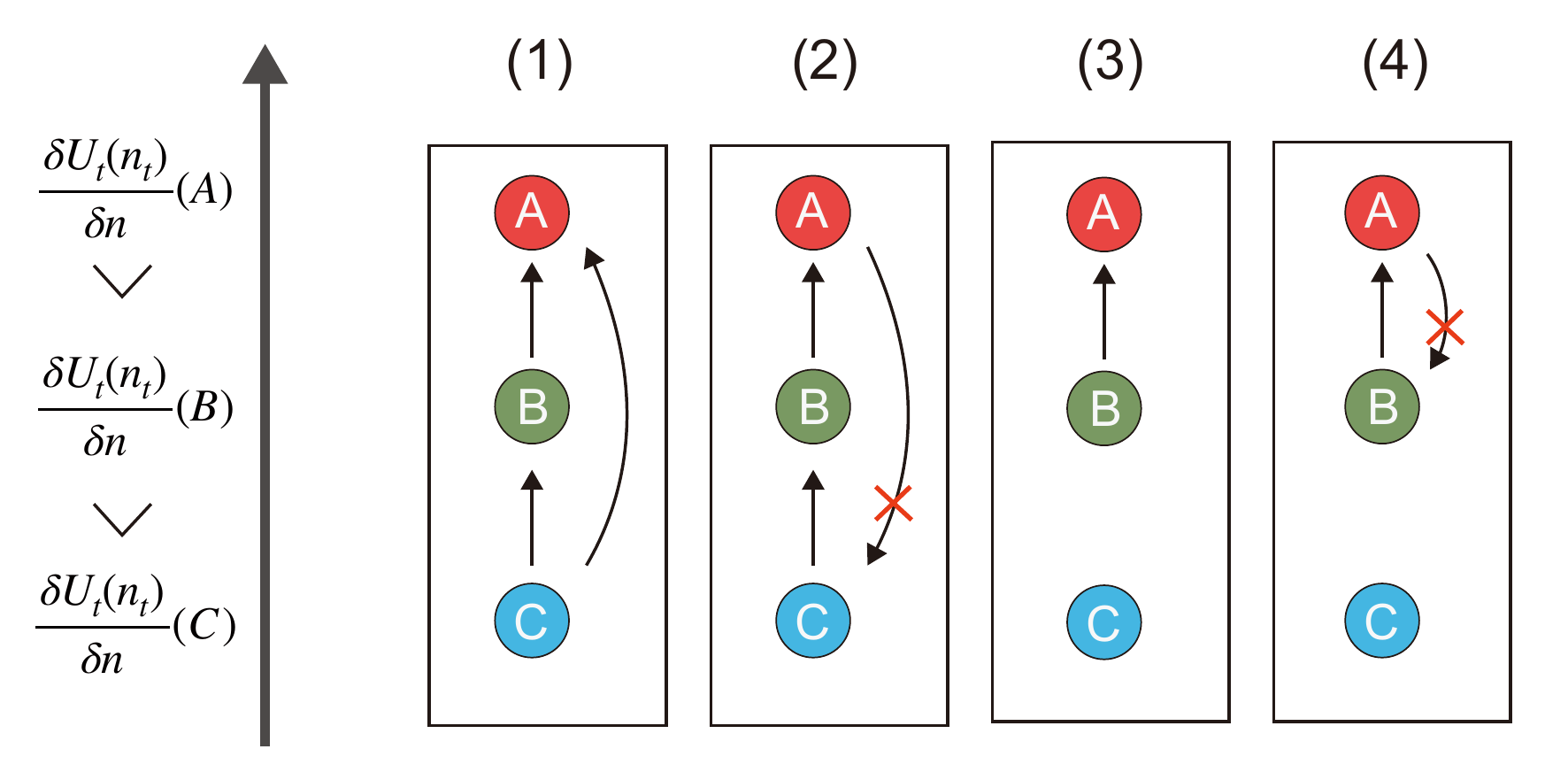}
\caption{Examples of type-switching graphs of three types $X=\{ A, B, C \}$. While acyclic graphs (1) and (3) are allowed in the gradient flow, graphs (2) and (4) contain a cycle and never appear in the gradient flow. For graph (2) and (4), one of the edges are incompatible with the ordering by $\frac{\delta U_t(n_t)}{\delta n}$}
\label{fig:switching_graph}
\end{figure}

\section{Emerging single-cell rules}

While our gradient flow is derived from the optimization at the population level, i.e., what kind of populational change is better than others, it also determines the behaviors of each cell, i.e., what a cell of type $x$ should do or should not do. We will show two such rules derived from the gradient flow structure. In both rules, the functional derivative $\frac{\delta U_t(n_t)}{\delta n}(x)$ of the utility function plays a vital role. One can interpret this function as a utility function at the single-cell level because it defines which type is preferable to the other types. It is in contrast to the original utility function $U_t(n)$, which defines a metric only for the population of cells.

\subsection{Unidirectional phenotypic transition}
In multicellular systems, unidirectional phenotypic switchings are commonly observed. For example, the T-cell immune response model does not have bidirectional or cyclic type switchings. We show that such unidirectionality is tightly linked to the gradient flow structure.

To this end, we consider a type-switching graph $G_t$. The nodes of the graph $G_t$ are types $X$, and an edge from type $x$ to $y$ exists if and only if the type-switching rate is not zero, i.e., $v_t^\dagger(x,y)>0$. One can show that this graph $G_t$ for any gradient flow is always acyclic (Fig.~\ref{fig:switching_graph}).
In the simplest case, the types $x$ and $y$ cannot have bidirectional type switching because $v_t^\dagger(x,y)$ and $v_t^\dagger(y,x)$ cannot be simultaneously positive from Eq.~\eqref{eq:optimal_switch_rate}.
Depending on the sign of $\overline{\nabla}\frac{\delta U_t(n_t)}{\delta n}(x,y)$, either or both of $v_t^\dagger(x,y)$ and $v_t^\dagger(y,x)$ is zero. In other words, cells of type $x$ switch to type $y$ if and only if $y$ is better than $x$, which means that
\begin{equation}
	\frac{\delta U_t(n_t)}{\delta n}(x) < \frac{\delta U_t(n_t)}{\delta n}(y).
\end{equation}
If we place all the types $X$ vertically in the order of $\frac{\delta U_t(n_t)}{\delta n}$ as in Fig.~\ref{fig:switching_graph}, every type-switching edge points upward.
Thus, more generally, three or more types cannot have any cyclic type switching. 
If a cycle exists, at least one of the edges points downward, which contradicts the ordering by $\frac{\delta U_t(n_t)}{\delta n}$.

One can view $\frac{\delta U_t(n_t)}{\delta n}$ as a kind of epigenetic landscape of single cells \cite{Waddington1957}. 
We note that it depends on time $t$ and the current population $n_t$, which could be interpreted as variations of the epigenetic landscape due to time-dependent external signals and cell-cell interactions. 


\subsection{Coupling}
Growth, immigration, and phenotypic switching in multicellular systems are not independent. For example, in the T-cell immune response model, the growth and type-switching rates change simultaneously when the environmental condition changes. We show that the gradient flow structure implies cooperative relationships among growth, immigration, and type-switching rates.

Consider the simplest setting where all the weights are finite and constant, $w_g(x)=w_m(x)=w_v(x,y)=1 ~\forall x,y\in X$. From the fact that the growth rates, immigration rates, and type-switching rates (Eqs.~\eqref{eq:optimal_rates}) have the same term $\frac{\delta U_t(n_t)}{\delta n}$, we find some relations among them.
When there is an immigration flux to type $x$, the growth rate of type $x$ must be positive and vice versa:
\begin{equation}\label{eq:immig_growth_coupling}
	m_t^\dagger(x) > 0 \Leftrightarrow g^\dagger_t(x) > 0.
\end{equation}
When the type-switching rate from type $x$ to $y$ is positive, the growth rates of the source type $x$ must be lower than the growth rate of the destination type $y$
\begin{equation}\label{eq:switch_growth_coupling}
	v_t^\dagger(x,y) > 0 \Leftrightarrow g_t^\dagger(x) < g_t^\dagger(y).
\end{equation}
One can intuitively understand these effects as cooperation among growth, immigration, and type-switching to achieve the same goal: maximizing the utility function.

The T-cell immune response model has this cooperative coupling as long as the inequalities in Eq.~\eqref{eq:ineq_params_Tcell} are satisfied.
The coupling between growth and type-switching was also predicted in Furusawa and Kaneko's model: the growth rate of stem-type cells is lower than differentiated-type cells \cite{Furusawa2001}.
Moreover, this kind of coupling has been observed in cell biology: cells undergoing differentiation stop the cell cycle and do not divide \cite{Ruijtenberg2016}.
Other coupling properties predicted from the gradient flow structure can be used to search for the structure in actual biological systems.

\section{Discussion}

Unveiling a potential gradient flow structure of a given population dynamics is reduced to identifying the utility and cost functions.
It would be desirable to have a systematic way of identifying these functions.
An approach is to infer the utility function from experimental data by using techniques in machine learning and single-cell omics \cite{AbAzar2020,Zhou2021}. 
Another approach is to derive the cost function from the physical and thermodynamic principles, e.g., by the large deviation theory \cite{Mielke2014}. 
In either case, our framework will serve as a basis for linking the single-cell and population properties.

Finally, it should be noted that a given population dynamics may not always fall into the class of gradient flow in the strict sense. Some modifications of the T-cell model (Eqs.~\eqref{eq:T-cell_model}–\eqref{eq:on_off_input}) can violate the conditions to be a gradient flow. 
Nevertheless, the gradient-flow-like behaviors can still be preserved if the modification is moderate, and the utility monotonically increases in time (see Appendix for more detail).
Thus, our theory can be used to search for such behaviors.
Moreover, we can further extend the notion of gradient flow \cite{Kraaij2020} to accommodate oscillatory components, e.g. cell cycle, and others. 
It expands the applicability of our approach to a wide range of multicellular phenomena and will be pursued.

\begin{acknowledgments}
We thank Kenji Itao for his helpful comments.
The first author is financially supported by the JSPS Research Fellowship Grant JP21J21415.
This research is supported by JST (JPMJCR2011, JPMJCR1927) and JSPS (19H05799).

\end{acknowledgments}

\bibliography{main}

\end{document}